\documentclass[prd,superscriptaddress,altaffilletter,showpacs,nofootinbib]{revtex4}
\usepackage{amssymb}
\usepackage{amsmath}
\usepackage{graphicx}
\usepackage{subfigure}
\usepackage{epsfig}
\usepackage{epstopdf}
\usepackage[usenames,dvipsnames]{color}
\usepackage{dcolumn}
\usepackage{bm}
\usepackage{color}

\begin{document}
\title{Universal spin-$1/2$ fermion field localization on a 5D braneworld }
\author{Nandinii Barbosa-Cendejas}\email{nandini@fis.unam.mx}
\affiliation{Instituto de Ciencias F\'{\i}sicas, Universidad Nacional Aut\'onoma de M\'exico,
Apdo. Postal 48-3, 62251 Cuernavaca, Morelos, M\'exico.}
\affiliation{Facultad de Ingenier\'ia El'ectrica, Universidad Michoacana de San Nicol\'as de Hidalgo,  
Morelia, Michoacan, M\'exico.}
\author{Dagoberto Malag\'on-Morej\'on}\email{malagon@fis.unam.mx}
\affiliation{Instituto de Ciencias F\'{\i}sicas, Universidad Nacional Aut\'onoma de M\'exico,
Apdo. Postal 48-3, 62251 Cuernavaca, Morelos, M\'exico.}
\author{Refugio Rigel Mora-Luna}\email{rigel@fis.unam.mx}
\affiliation{Instituto de Ciencias F\'{\i}sicas, Universidad Nacional Aut\'onoma de M\'exico,
Apdo. Postal 48-3, 62251 Cuernavaca, Morelos, M\'exico.}

\date{\today}
\begin{abstract}
In this work we present a refined method for the localization of
spin--$\frac{1}{2}$ fermions on the 5D braneworld paradigm. We begin
by proposing a more natural ansatz for the Yukawa coupling in the 5D
bulk fermionic action, that guarantees the localization of the
ground states for the 4D fermions with right-- or left-- chirality.
In earlier works the existing freedom on the form of the Yukawa
coupling was used in a rather speculative way depending on the type
of model, the  ansatz proposed in this work is suitable for thin and
thick braneworld models and can be applied to branes made of a
scalar field or not and in this sense it is the more natural choice.
Furthermore, we show that the fermion ground states localization
allow us to show the absence of tachyonic modes in the left-- and
right-- chiral Kaluza-Klein mass spectrum. More precisely, we show
that localization of gravity in the 5D braneworld implies the
localization of the spin--$\frac{1}{2}$ fermions. 
\end{abstract}

\maketitle
\section{Introduction}
At the end of the last century a new approach for the hierarchy
problem in high energy physics came into light, it was a geometric
point of view born from the generalization of the ideas of Kaluza
and Klein. The main idea was to use a warped ansatz for the metric
that could in principle depend on the extra dimensions in order to
solve the mass hierarchy problem. Some of the models that raised
this idea were the thin braneworld models
\cite{RubakovPLB1983136}-\cite{rs}, which prove to successfully
address the mass hierarchy problem and the gravity localization in a
$3+1$-braneworld. This scenario of the universe located at a
space-time submanifold was pointed out in string theory
\cite{HOWI1}-\cite{HOWI2}. An important issue of this scenario was
the mechanisms for the localization of matter fields in this
submanifold called thin brane. Nowadays, a more general picture
arises, the thick braneworld models \cite{deWolfe}-\cite{HRMGR},
which offer a more realistic approach because the submanifold is not
located at a singularity as in the thin brane models,  and in some
cases the effects of the thickness can be important. This fact can
lead to a more phenomenological study of the physics on the brane
because the matter is a non-singular mass distribution localized
along the fifth dimension. Furthermore the thick brane models may
also have a more fundamental motivation originating from
supergravity theories, as shown in \cite{Tevfik}. In this context any
conjecture about the mechanism for the matter field confinement in
the brane can be made \cite{BajcPLB2000}-\cite{HQuChu}.
However the assumptions can be very speculative, with the effect of matter localization
mechanisms preferring some models to others.  In fact, the method for localizing the matter appears to be a
discriminant among the braneworld models, for example spin--$0$, the
scalar field localization performed in
\cite{BajcPLB2000}-\cite{Guoetal2013}, is closely related to the
localization of gravity; in other words, the localization of gravity
guarantees the localization of the KK ground state for the spin--$0$
matter field. A drawback on the other hand, is that the localization
mechanism performed by \cite{DvaliShifman}-\cite{OdaPLB2000113} was
not suitable to localize gauge fields with spin--$1$  for some thick and
thin branewords. However a recent work \cite{XYYI} menaced to
localize the  zero mode of the gauge fields with spin--$1$ in the 5D
braneworld models, where the localization had not been suitable.

The localization mechanism employed for matter fields living in a 5D
braneworld scenarios has been the subject of many studies, yet the
most popular method for the localization of spin--$\frac{1}{2}$
fermions is formulated  in a rather speculative way
\cite{MelfoPRD2006}-\cite{HQuChu}. This is because of the freedom
one has to propose the Yukawa coupling term. The fermion
localization can be performed in models of thin and thick branes
arising from a scalar field $\phi$ (like the ones in
\cite{Barbosa-Cendejas:2013cxa}-\cite{mgio1}), where we can choose
the interaction as $\eta\mathcal{F}(\phi(z))\bar\Psi \Psi$, imposing
as a necessary requirement that the arbitrary function
$\mathcal{F}(\phi(z))$ is an odd function in the fifth dimension
$z$. Otherwise we should change the localization mechanism by
introducing a new form for the Yukawa coupling that is $\eta\bar\Psi
\Gamma^{N}\partial_{N}\mathcal{F}(\phi(z))\gamma^{5}\Psi$ as shown
in \cite{YXuFWei}. In the case of branes that are not made of a
scalar field $\phi(z)$, the proposal for the Yukawa coupling
$M\mathcal{F}(z)\bar\Psi \Psi$ arises as a mass distribution
$M\mathcal{F}(z)$ in the 5D bulk, where the field $\mathcal{F}(z)$
must be an odd function of $z$ (see for instance \cite{Guoetal2013}
and \cite{Oda:2001ss}). If $\mathcal{F}(z)$ fails to be an odd
function, the localization mechanism must be changed by proposing
the Yukawa coupling as $M\bar\Psi
\Gamma^{N}\partial_{N}\mathcal{F}(z)\gamma^{5}\Psi$. In the present
letter we find a more natural ansatz for the Yukawa coupling term
$F(z)\bar\Psi\Psi$ added to the 5D bulk action. This emerges from
the localization mechanism where $F(z)$ inherits its odd nature
directly from the geometry shape of the warp factor $e^{A(z)}$, as
will be shown later. Moreover the coupling $F(z)\bar\Psi\Psi$ avoids
arbitrariness as it is independent of the braneworld model and
guarantees the localization for the ground states of the KK spectrum
for the right-- or left-- chiral spin--$1/2$ fermions. This choice
also leaves open the possibility of finding more bound states since
we can manipulate the depth of the 5D KK potentials, always looking
to avoid tachyonic states in the KK spectrum. We should note that
only in some cases when the brane is made of a scalar field
$\phi(z)$, it is possible to find a bijection between $\phi$ and
$z$. For this cases the interaction $F(z)\bar\Psi\Psi$, can be
written as a function that depends on $\phi$ only through
$F(\phi(z))\bar\Psi \Psi$. Thus, our proposal for $F(z)$ in the
mechanism of localization give us the choice of $F(\phi(z))$. In
this context, a Yukawa coupling term $F(z)\bar\Psi\Psi$ that can be
suited for thin and thick branes arising from a scalar field or not
is the most natural choice.

This work is presented in three sections; we first give a
brief introduction to the mechanism of
spin--$1/2$ fermions localization, in Section three we address the
choice for the $F(z)$ function with allows us to obtain a left-- and
right-- chiral KK mass spectrum free on tachyonic instabilities, as
well as the localization of the ground sate for the fermions with
left-- or right-- chirality. Furthermore in Section four we show
that in our natural proposal the localization of gravity on the
braneworld implies the localization of spin--$1/2$ fermions as well;
finally we present our conclusions in Section five.

\section{ localization of Spin--$1/2$ fermion fields on a braneworld}
The method we use is the standard mechanism for the localization of
spin--$1/2$ fermions, see for example
\cite{MelfoPRD2006}--\cite{HQuChu}, the procedure is shown in the
following lines. The ansatz for the 5D geometry is
\begin{equation}
\label{conformmetric}
 dS_{5}^{2}=e^{2A(z)} \left(\hat g_{\mu\nu}(x)dx^{\mu}dx^{\nu}
 +dz^{2}\right),
\end{equation}
where $z$ is the 5D coordinate, $\hat g_{\mu\nu}$ is the 4D induced
metric on the brane with $\mu,\nu=0,1,2,3$ and the function $e^{A}$
is the warp factor. In this models we can assume, as it is shown in
\cite{0910.0363}--\cite{thickreview}, that gravity is localized in
the brane, meaning  that the zero mode of the KK spectrum associated
with the 4D graviton $\Psi_{grav_{0}}\sim e^{\frac{3}{2}A(z)}$ is a
normalizable function on the fifth dimension.  We suppose that the
5D fermionic fields have a weak interaction with gravity (we do not
consider the  backreaction effects) in such a way that the
localization properties of the 5D graviton $\Psi_{grav_{0}}$ do
not change.

 In 5D spacetime, fermions are
four--component spinors and their Dirac structure can be described
by $\Gamma^M= e^M_{~\bar{M}} \Gamma^{\bar{M}}$ with $e^M_{~\bar{M}}$
being the viervein and $\{\Gamma^M,\Gamma^N\}=2g^{MN}$. In this
Subsection, $\bar{M}, \bar{N}, \cdots =0,1,2,3,5$ and $\bar{\mu},
\bar{\nu}, \cdots =0,1,2,3$ denote the 5D and 4D local Lorentz
indices, respectively, and $\Gamma^{\bar{M}}$ are the gamma matrices
in 5D flat spacetime.

For our set-up, the viervein and the gamma matrices are defined
trough
\begin{eqnarray}
e_M ^{~~\bar{M}}= \left(%
\begin{array}{ccc}
  e^{A} \hat{e}_\mu^{~\bar{\nu}} & 0  \\
  0 & e^{A}  \\
\end{array}%
\right),
\qquad\qquad  \Gamma^M=e^{-A}(\hat{e}^{\mu}_{~\bar{\nu}}
 \gamma^{\bar{\nu}},-i\gamma^5)=e^{-A}(\gamma^{\mu},-i\gamma^5),
\label{vielbein_e}\nonumber
\end{eqnarray}
where $\gamma^{\mu}=\hat{e}^{\mu}_{~\bar{\nu}}\gamma^{\bar{\nu}}$,
$\gamma^{\bar{\nu}}$ and $\gamma^5$ are the usual flat gamma
matrices in the 4D Dirac representation.

The Dirac action of a spin--1/2 fermion with a mass term can be expressed as
\cite{Oda}
\begin{eqnarray}
S_{\frac{1}{2}} = \int d^5 x \sqrt{-g} \left[\bar{\Psi}i\Gamma^M
          \left(\partial_M+\omega_M\right) \Psi
          -  F(z) \bar{\Psi}\Psi\right]. \label{DiracAction}
\end{eqnarray}
Here $\omega_M$ is the spin connection defined as $\omega_M=
\frac{1}{4} \omega_M^{\bar{M} \bar{N}} \Gamma_{\bar{M}}
\Gamma_{\bar{N}}$ while $ \omega_M ^{\bar{M} \bar{N}}$ is defined as
\begin{eqnarray}
 \omega_M ^{\bar{M} \bar{N}}
   &=& \frac{1}{2} {e}^{N \bar{M}}\left(\partial_M e_N^{~\bar{N}}
                      - \partial_N e_M^{~\bar{N}}\right)
    - \frac{1}{2} {e}^{N\bar{N}}\left(\partial_M e_N^{~\bar{M}}
                      - \partial_N e_M^{~\bar{M}}\right)  \nonumber \\
   && - \frac{1}{2} {e}^{P \bar{M}} {e}^{Q \bar{N}}\left(\partial_P e_{Q
{\bar{R}}} - \partial_Q e_{P {\bar{R}}}\right) {e}_M^{~\bar{R}},
\end{eqnarray}
and $F(z)$ is some general scalar function of the extra dimensional
coordinate $z$. We will discuss about the properties of the scalar
function $F(z)$ later in Section three, in the context of the
localization of KK fermion modes. The non--vanishing components of
the spin connection $\omega_M$ for the background metric
(\ref{conformmetric}) are
\begin{eqnarray}
  \omega_{M} =\frac{1}{2}(\partial_{z}A) \gamma_\mu \gamma_5
             +\hat{\omega}_\mu, \label{spinConnection}
\end{eqnarray}
where $\hat{\omega}_\mu=\frac{1}{4} \bar\omega_\mu^{\bar{\mu}
\bar{\nu}} \Gamma_{\bar{\mu}} \Gamma_{\bar{\nu}}$ is the spin
connection derived from the metric
$\hat{g}_{\mu\nu}(x)=\hat{e}_{\mu}^{~\bar{\mu}}(x)
\hat{e}_{\nu}^{~\bar{\nu}}(x)\eta_{\bar{\mu}\bar{\nu}}$. Thus, the
equation of motion corresponding to the action (\ref{DiracAction})
varying over $\bar \Psi$ can be written as
\begin{eqnarray}
 \left[ i\gamma^{\mu}(\partial_{\mu}+\hat{\omega}_\mu)
         + \gamma^5 \left(\partial_z  +2 \partial_{z} A \right)
         -e^A F(z)
 \right ] \Psi =0, \label{DiracEq1}
\end{eqnarray}
where $\gamma^{\mu}(\partial_{\mu}+\hat{\omega}_\mu)$ is the Dirac
operator on the brane.

We turn now to investigate the 5D Dirac equation (\ref{DiracEq1}),
and write the spinor in terms of 4D effective fields. On account of
the fifth gamma matrix $\gamma^{5}$, we anticipate that the left--
and right--handed projections of the 4D part behave in a different
manner. From Eq. (\ref{DiracEq1}), we propose the following KK
decomposition for the 5D spinors $\bar\Psi$ and $\Psi$
\begin{equation}
 \Psi= e^{-2A}\left(\sum_n\psi_{Ln}(x) L_n(z)
 +\sum_n\psi_{Rn}(x) R_n(z)\right).
\end{equation}
The KK decomposition proposal for the 5D Dirac
spinors is inspired in the chirality proyectors, this decomposition
imposes chirality on the massive KK modes inherited from the 4D
projectors
\begin{eqnarray}
\psi_{Ln}&=&\frac{1}{2}\left(I_{4}-\gamma^{5}\right)\psi(x)_{n},\\
\psi_{Rn}&=&\frac{1}{2}\left(I_{4}+\gamma^{5}\right)\psi(x)_{n},
\end{eqnarray}
where
$\gamma^{5}\psi_{Ln}=\frac{1}{2}\left(\gamma^{5}-I_{4}\right)\psi(x)_{n}=-\psi_{Ln}$
and
$\gamma^{5}\psi_{Rn}=\frac{1}{2}\left(\gamma^{5}+I_{4}\right)\psi(x)_{n}=\psi_{Rn}$
are the left-handed and right-handed components of a 4D Dirac field,
respectively. In order to decouple the  Dirac equation
(\ref{DiracEq1}) in its  4D and  5D part, we must assume that the
left-- and the right--handed eigenfunctions $\psi_{Ln}(x)$ and
$\psi_{Rn}(x)$ satisfy the 4D Dirac equations. The consequences of
this assumption allow us to write a system of two coupled
differential equations for the eigenvalues of the KK modes
$L_{n}(z)$ and $R_{n}(z)$  given by:
\begin{eqnarray}
 \left(\partial_z
                  + e^A F(z) \right)L_n(z)
  &=&  ~~m_n R_n(z), \label{CoupleEq1a}  \\
 \left(\partial_z
                  - e^A F(z) \right)R_n(z)
  &=&  - m_n L_n(z). \label{CoupleEq1b}
\label{CoupleEq1}
\end{eqnarray}
The system above can be decoupled by mixing  (\ref{CoupleEq1a}) and (\ref{CoupleEq1b}) equations and after some algebra we get a pair of  Schr\"odinger type equations
\begin{eqnarray}
  \Big(-\partial^2_z + V_L(z) \Big)L_n
            &=&m_{n}^{2} L_n,~~
       \label{SchEqLeftFermion}  \\
  \Big(-\partial^2_z + V_R(z) \Big)R_n
            &=&m_{n}^{2} R_n,
   \label{SchEqRightFermion}
\end{eqnarray}
where the corresponding left and right potentials read
\begin{eqnarray}
  V_L(z)&=& \left(e^{A(z)}F(z)\right)^{2}
     - \left(e^{A(z)}F(z)\right)', \label{VL}\\
  V_R(w)&=&   \left(e^{A(z)}F(z)\right)^{2}
     + \left(e^{A(z)}F(z)\right)'. \label{VR}
\label{Vfermion}
\end{eqnarray}
In this notation a prime denotes  $\partial_{z}$.

\section{The proposal for $F(z)$}
We can immediately see that the equations (\ref{SchEqLeftFermion})
and (\ref{SchEqRightFermion}) equations are restricted by the
behavior of the function $F(z)$. In what follows we will find a way
to choose this function by directly looking at the integrability
conditions on $L(z)$ o $R(z)$ in equations
(\ref{SchEqLeftFermion}) and (\ref{SchEqRightFermion}). We are
mainly interested in the localization properties  of the massless
mode of the KK excitations. Let us thus consider the $m_{n}=0$ case;
if we take $y(z)=e^{A(z)}F(z)$, we can rewrite the potential $V_{L}(z)$ as
\begin{equation}
  V_L(z)= y^{2}
     - y'. \label{VLy}
\end{equation}
The last equation is a Riccati differential equation and a particular solution can be
obtained by performing the change of variable  $y=-\frac{{\it Z'}}{{\it Z}}$. Then
we have a differential equation for the new variable that reads
\begin{equation}\label{Zz}
-{\it Z''(z)}+V_{L}(z){\it Z(z)}=0.
\end{equation}
By taking a closer look at equation (\ref{Zz}) we see that it is
a Schr\"odinger--like equation, the same equation that the massless
version of (\ref{SchEqLeftFermion}). Thus, $\it{Z}=L_{0}$ is a
natural choice for its solution. On the
other hand, as we mentioned above, the wave function of the massless
graviton is normalized and takes the form $\Psi_{grav_{0}}\sim
e^{\frac{3}{2}A(z)}$ (see \cite{Csaki_NPB_2000}-- \cite{HRMGR}).
This fact suggests that a simple choice for  $L_{0}={\it Z}$ can be
a power law of the zero mode for the graviton, namely
\begin{equation}
{\it Z}(z)=e^{MA(z)},
\label{zansatz}
\end{equation}
where $M$ is a real and positive constant, that can be set in such a
way that the ${\it Z}(z)$ function is normalizable in the fifth
dimension. By taking into account the above relation, the coupling
$F(z)$ can be written as
\begin{equation}\label{fofz}
F(z)=M\partial_{z}{e}^{-A(z)}.
\end{equation}
The above relations imply that the Yukawa coupling  is entirely
determined by the bulk geometry. In references \cite{mgio,mgio1}, the
authors studied braneworld models for a class of geometries where
the warp factor has the following behaviour at infinity
\begin{equation}
e^{A} \sim \frac{1}{|z|^{\gamma}}, \quad \mbox{when} \quad z
\rightarrow \infty.
\label{eq:asimpwarpfactor}
\end{equation}
 In the cited references it was shown that if the 4D Planck mass is
finite, a massless graviton is localized  and there are no not space--time
singularities, and the values adopted by the exponent are defined for
$1/3 < \gamma \leq 1.$ By considering
this class of metrics, the coupling $F$ takes the form
$$
F \sim M \,\gamma \,\mbox{sgn}(z) |z|^{\gamma -1}, \quad
\mbox{when}\quad z\rightarrow \infty.
$$
Hence, the field $F$ is finite along the extra dimension. This fact
is important because it guarantees the stability of the field $F$.

Now we take the ansatz for $F(z)$ and substitute it into the
equations (\ref{CoupleEq1a}) and (\ref{CoupleEq1b}). This allows us to
recast the left and right pair of equations as follows
\begin{eqnarray}
 \left(\partial_z
                  - M A^{'} \right)L_n(z)
  &=&  ~~m_n R_n(z), \label{CoupleEq1a2}  \\
 \left(\partial_z
                  + M A^{'} \right)R_n(z)
  &=&  - m_n L_n(z). \label{CoupleEq1b2}
\end{eqnarray}
In order to obtain the zero modes for the fermions with right-- and
left--chirality we must set  $m_n=0$  in (\ref{CoupleEq1a2}) and
(\ref{CoupleEq1b2}) to obtain:
\begin{eqnarray}
  L_0&\propto & e^{MA(z)}, \label{zerol}\\
  R_0&\propto & e^{-MA(z)}. \label{zeror}
\end{eqnarray}

The normalization of the zero mode spin-2 graviton ensures that
$e^{A} \rightarrow 0$ when $z\rightarrow \pm \infty.$  Then, for a
given sign of $M$ the above relations tell us that it is not
possible to have both massless left-- and right--chiral KK fermion
modes localized on the brane at once, since when one is localized,
the other one is not.

 In general if we  ask for
$e^{MA}$ to be a normalizable function on the fifth dimension
and combine the equations (\ref{CoupleEq1a2}) and
(\ref{CoupleEq1b2}), we can obtain the Schr\"{o}dinger--like
equations for the left-- and right--chiral KK modes of fermions:
\begin{eqnarray}
  -\partial^2_z L_{n} + V_{L}L_n
            &=&m_{L_n}^{2} L_n,~~
   \label{SchEqLeftFermionn2}  \\
  -\partial^2_z R_{n} + V_{R}R_n
            &=&m_{R_n}^{2} R_n.
   \label{SchEqRightFermionn2}
\end{eqnarray}
Then the ansatz for  $F(z)$ fixes the form of the potentials $V_{L,R}$ that can be
recast as follows
\begin{eqnarray} \label{LANDRPOT1}
  V_{L}(z)&=& M^{2}A'^{2}+MA'' ,   \\
  V_{R}(z)&=& M^2A'^{2}-MA''.
   \label{LANDRPOT2}
\end{eqnarray}
By analyzing the shape of the potentials (\ref{LANDRPOT1}) and
(\ref{LANDRPOT2}), it is easy to see that the depth of the potential
wells $V_{L,R}(z)$ is determined by the size of the parameter $M$.
Furthermore, if we demand that the fermions with right/left chirality
are localized in the 3-brane and that they present a symmetric
distribution around the brane origin, then the $Z(z)$ must
necessarily be an even function, implying that
$F(z)=M\partial_{z}{e}^{-A(z)}$, which is an odd function. This
statement can be explicitly demostrated by substituting $F(z)$ into
equations (\ref{VL}) and (\ref{VR}). If we assume that
$e^{A}$ is an even function, then the potentials $V_{L,R}$ are also
even functions around the center of the brane in the fifth
dimension.

We show that the KK mass spectrum for the fermions with
right-- and left--chirality is free of tachyonic modes (see
\cite{A.Andrianov} and references there in) by writing their quantum
mechanic supersymmetric analogue form from the Schr\"odinger--like
equations (\ref{SchEqLeftFermionn2}) and (\ref{SchEqRightFermionn2})
\begin{eqnarray}
  \left(-\partial_z -MA'\right)\left(\partial_z -MA' \right)L_{n}
            &=&m_{L_n}^{2} L_n,~~
   \label{FactorSchEqLeftFermion2}  \\
  \left(\partial_z -MA'\right)\left(-\partial_z -MA' \right)R_{n}
            &=&m_{R_n}^{2} R_n.
   \label{FactorSchEqRightFermion}
\end{eqnarray}
In contrast with the zero mode, it is difficult to obtain analytical
solutions for the massive spectrum. There is no dictated procedure
to obtain the eigenfunctions with non-zero mass. However, in our
case we can extract some qualitative information about the spectrum.
For the class of geometries described in equation (\ref{DiracEq1})
the left potential have two possible behaviors at infinity, the
first one  is when $V_{L}>0$ as $z \rightarrow \pm \infty $. Clearly
in this scenario there is a mass gap between the bound states and
the continuum states. If we look at equations
(\ref{eq:asimpwarpfactor}) and (\ref{zerol}) it easy to see that for
the second case the potential at infinity behave as
\begin{equation}\label{behaviorVL}
V_{L} \sim \frac{M \gamma}{z^{2}}\left(M\gamma+1\right).
\end{equation}
In \cite{Csaki_NPB_2000} it is shown that this kind of potential
produces a continuous massive spectrum and  that the  continuum
states are decoupled from the ground state. We can also see from the
potentials (\ref{LANDRPOT1}) and (\ref{LANDRPOT2}) that $V_{R}$ must
have a similar behavior as the $V_{L}>0$ as $z\rightarrow \pm \infty
$, because $A''\rightarrow0$ when $z\rightarrow\pm\infty$. Moreover
when $z\rightarrow0$, the right potential is restricted by the
inequality $V_{R}>V_{L}$.  Taking in to account the above arguments,
we  can conclude that there is a mass gap between the bound states and
a the tower of continuum states. The only difference between the KK
mass spectra for $L_{m}$ and $R_{m}$, is that the KK mass spectrum
for $R_{m}$ has less bound states than its analogue $L_{m}$.
This fact can be seen more clearly because both ground states
$L_{0}$ and $R_{0}$ cannot be localized at the same time, this
effect is associated to the depth of the potentials, that in $V_{R}$
is more suppressed by the second term on the right side of the
equation (\ref{LANDRPOT2}). Otherwise, when $V_{L}<0$ at $z=0$, and
its behavior at infinity is again described by equation (\ref{behaviorVL}), then
the shape of the right  potential turns out to be $V_{R}> 0$ when
$z\rightarrow 0$, and the asymptotic behavior is $V_{R}\rightarrow0$
when $z \rightarrow \infty$. For this case there are no localized
bound states.

\section{Relation between gravity and fermion localization}
A more direct relation between gravity localization and  spin-$1/2$
fermionic fields localization can be made if we set
$M\geq\frac{3}{2}$. In particular if we choose $M=3/2$, a direct
relation between the localization of the left chiral ground state
and the zero mode of the KK spectrum for the 5D graviton
$\Psi_{grav_{0}}\sim e^{\frac{3}{2}A(z)}$ can be made.  The relation
is immediate if we see the expresions for the potentials
(\ref{LANDRPOT1}) and (\ref{LANDRPOT2}) and the analogue
supersymmetric quantum mechanic potential for the KK graviton ground
state equation \cite{Csaki_NPB_2000},
\begin{eqnarray} \label{LANDRPOT13}
  V_{L}(z)&=& \frac{9}{4} A'^{2}+\frac{3}{2} A'' ,   \\
  V_{R}(z)&=& \frac{9}{4}A'^{2}-\frac{3}{2}A'' .
   \label{LANDRPOT23}
\end{eqnarray}
It is important to state that if we set $M=-M$ in the equations
(\ref{SchEqLeftFermionn2}) and (\ref{SchEqRightFermionn2}), a
similar treatment can be performed. The results of this new scenario
are the following: the zero mode of the right chirality fermion
$R_{0}$ is localized, while its analogue left ground state $L_{0}$
remains delocalized. The analogous supersymmetric quantum mechanic form
of the Schr\"odinger equations (\ref{FactorSchEqLeftFermion2}) and
(\ref{FactorSchEqRightFermion}) can be written as
\begin{eqnarray}
  \left(-\partial_z -MA'\right)\left(\partial_z -MA' \right)R_{n}
            &=&m_{R_n}^{2} R_n,~~
   \label{FactorSchEqLeftFermion22}  \\
  \left(\partial_z -MA'\right)\left(-\partial_z -MA' \right)L_{n}
            &=&m_{L_n}^{2} L_n.
   \label{FactorSchEqRightFermion22}
\end{eqnarray}
Once we obtain the eigenfunctions  $L_{n}$, $R_{n}$ associated
 to the bound sates of the KK spectrum via the Schr\"odinger--like equation, we can perform a dimensional
 reduction over the 5D action and use them along with the determinant of the metric $\sqrt{-g}$ and the warp factor
$e^{A}$ to derive the corresponding field configuration in 4D
\begin{eqnarray}
 S_{\frac{1}{2}} &=& \int d^5 x \sqrt{-g} ~\bar{\Psi}
     \left[ i\Gamma^M (\partial_M+\omega_M)
     -F(z)\right] \Psi,  \nonumber \\
  &=&\sum_{n}\int d^4 x \sqrt{-\hat{g}}
    ~\bar{\psi}_{n}
      \left[i\gamma^{\mu}(\partial_{\mu}+\hat{\omega}_\mu)
        -m_{n}\right]\psi_{n},~~~
\end{eqnarray}
where the eigenfunctions $L_{n}$ and $R_{n}$ form a closed set of
linearly independent functions that satisfy the following
orthogonality relations:
\begin{eqnarray}
 \int_{-\infty}^{+\infty} L_m L_ndz
   &=& \delta_{mn}, \label{orthonormalityFermionL} \\
 \int_{-\infty}^{+\infty} R_m R_ndz
   &=& \delta_{mn}, \label{orthonormalityFermionR}\\
 \int_{-\infty}^{+\infty} L_m R_ndz
   &=& 0. \label{orthonormalityFermionLR}
\end{eqnarray}

\section{Conclusions}
In this letter we have proposed a natural way of determine the
Yukawa coupling $F(z)\bar\Psi \Psi$ for the 5D bulk action, the form
of the $F(z)$ function allows us to localize the spin--$\frac{1}{2}$
fermionic fields in the paradigm of the 5D braneworld scenarios. The
structure acquired by the $V_{L,R}$ potentials is induced by the
ansatz for the $F(z)$ function and we show that this choice allows
the localization of the left-- or right--chiral KK zero modes
$L_{0}(z)$, $R_{0}(z)$ and excludes the tachyonic modes of the KK
mass spectrum $m_{L_{n}}$ and $m_{R_{n}}$. The proposal for the
function $F(z)$ leaves behind other speculative proposals since it
is independent of the braneworld model, hence the form for the
Yukawa coupling here presented is the most natural choice. This is
shown most evidently if we set $M=\pm 3/2$, which is a more
phenomenologically viable ansatz. This choice allows for a direct
relation between the localization of the ground state for the left--
or right--chiral KK mass spectrum and the localization of the zero
mode for the 4D graviton in the brane. As a bonus, we include a brief 
discussion  about the existence of a left-- or right--chiral KK zero mode 
decoupled from the corresponding continuous KK mass spectrum, as well as the
possible existence of a mass gap, due to the asymptotic nature of
the potentials $V_{R,L}$.

\begin{acknowledgements}
We would like to thank Dr. Carlos Hidalgo for the helpful 
comments and suggestions. RRML acknowledges a postdoctoral grant from CONACyT at ICF-UNAM, NBC
and DMM acknowledges a postdoctoral grant from DGAPA at ICF-UNAM, as
well as to ``Programa de Apoyo a Proyectos de Investigaci\'on e
Innovaci\'on Tecnol\'ogica'' (PAPIIT) UNAM, IN103413-3, Teor\'{i}as
de Kaluza--Klein, inflaci\'on y perturbaciones gravitacionales. All
authors thank SNI for support.
\end{acknowledgements}


\begin{thebibliography}{99}

\bibitem{RubakovPLB1983136}
V.A. Rubakov and M.E. Shaposhnikov,
{\it Do we live inside a domain wall?},
Phys. Lett.  B {\bf 125} (1983) 136.

\bibitem{Randjbar-DaemiPLB1986}
S. Randjbar-Daemi and C. Wetterich,
 {\it Kaluza-Klein solutions with noncompact internal spaces},
Phys. Lett.  B {\bf 166} (1986) 65.

\bibitem{DvaliShifman}
G. Dvali and M.A. Shifman,
{\it Domain walls in strongly coupled theories},
Phys. Lett. B {\bf 396} (1997) 64.

\bibitem{AntoniadisPLB1990}
 I. Antoniadis,
{\it A possible new dimension at a few Tev},
Phys. Lett. B {\bf 246} (1990) 377.

\bibitem{ADD}
 N. Arkani-Hamed, S. Dimopoulos and G. Dvali,
 {\it The hierarchy problem and new dimensions at a millimeter},
 Phys. Lett.  B {\bf 429} (1998) 263.

 \bibitem{AntoniadisArkani}
 I. Antoniadis, N. Arkani-Hamed, S. Dimopoulos and G. Dvali,
 {\it New dimensions at a millimeter to a Fermi and superstrings at a TeV},
 Phys. Lett.  B {\bf 436} (1998) 257.

\bibitem{rs1}
 L. Randall and R. Sundrum,
 {\it A Large Mass Hierarchy from a Small Extra Dimension},
 Phys. Rev. Lett. {\bf83} (1999) 3370.

\bibitem{rs}
 L. Randall and R. Sundrum,
 {\it An alternative to compactification},
 Phys. Rev. Lett. {\bf83} (1999) 4690.

 \bibitem{HOWI1}
P. Horava and E. Witten,
{\it Heterotic and Type I string dynamics from eleven dimensions},
Nucl. Phys. B {\bf 460} (1996) 506.

\bibitem{WI1}
E. Witten,
 {\it Strong coupling expansion of Calabi-Yau compactification},
Nucl. Phys. B {\bf 471} (1996) 135.

\bibitem{HOWI2}
P. Horava and E. Witten,
{\it Eleven-dimensional supergravity on a manifold with boundary},
Nucl. Phys. B {\bf 475} (1996) 94.

\bibitem{Neupane:2009ws}
I.~P.~Neupane,
{\it Extra dimensions, warped compactifications and cosmic acceleration},
Phys.\ Lett.\ B {\bf 683}, 88 (2010).
\bibitem{deWolfe}
O. DeWolfe, D.Z. Freedman, S.S. Gubser and A. Karch,
{\it Modeling the fifth dimension with scalars and gravity},
Phys. Rev. D {\bf 62} (2000) 046008.

\bibitem{Csaki_NPB_2000}
 C. Csaki, J. Erlich, T. Hollowood and Y. Shirman,
{\it Universal Aspects of gravity localized on thick branes},
Nucl. Phys. B {\bf 581} (2000) 309.

\bibitem{Barbosa-Cendejas:2013cxa}
 N.~Barbosa-Cendejas, A.~Herrera-Aguilar, U.~Nucamendi, I.~Quiros and K.~Kanakoglou,
{\it``Mass hierarchy, mass gap and corrections to Newton's law on thick branes with Poincare symmetry,''}
Gen.\ Rel.\ Grav.\  {\bf 46}, 1631 (2014).

\bibitem{0910.0363}
 A. Herrera-Aguilar, D. Malag\'on-Morej\'on, R.R. Mora-Luna and U. Nucamendi,
 {\it Aspects of thick brane worlds: 4D gravity localization,
 {smoothness, and mass gap}}.
 Mod. Phys. Lett. A {\bf 25} (2010) 2089.

\bibitem{thickreview}
V. Dzhunushaliev, V. Folomeev and M. Minamitsuji,
{\it Thick brane solutions}.
Rept. Prog. Phys. {\bf 73} (2010) 066901.

\bibitem{mgio}
M. Giovannini, {\it Localization of metric fluctuations on scalar branes},
Phys. Rev. D{\bf 65} (2002) 064008.

\bibitem{mgio1}
M. Giovannini, {\it Thick branes and Gauss--Bonnet self--interactions},
Phys.Rev. D {\bf 64} (2001) 124004.

\bibitem{Neupane:2009br}
 I.~P.~Neupane,
 {\it Accelerating universe from warped extra dimensions},
 Class.\ Quant.\ Grav.\  {\bf 26}, 195008 (2009).

\bibitem{1009.1684}
 A. Herrera-Aguilar, D. Malag\'on-Morej\'on and R.R. Mora-Luna,
  {\it Localization of gravity on a thick braneworld without scalar fields},
JHEP {\bf 1011} (2010) 015.

\bibitem{HRMGR}
G. Germ\'an, A. Herrera-Aguilar, D. Malag\'on-Morej\'on and R.R. Mora-Luna, R. da
Rocha, {\it A de Sitter tachyon thick braneworld},
JCAP {\bf2013} (2013) 035.

\bibitem{Tevfik}
N. Tevfik-Yilmaz,
{\it Supergravity induced interactions on thick branes},
Chin. Phys. B {\bf 23} (2014) 040401.

\bibitem{BajcPLB2000}
B. Bajc and G. Gabadadze,
{\it Localization of matter and cosmological constant on a brane in anti de Sitter space},
Phys. Lett. B {\bf 474} (2000) 282.

\bibitem{NonLocalizedFermion}
Y. Grossman and N. Neubert,
 {\it Neutrino masses and mixings in non-factorizable geometry},
Phys. Lett. B {\bf 474} (2000) 361.

\bibitem{Liu0708}
 Y.-X. Liu, X.-H. Zhang, L.-D. Zhang and Y.-S. Duan,
 {\it Localization of Matters on Pure Geometrical Thick Branes},
JHEP {\bf 0802} (2008) 067.

\bibitem{Liu0808}
Y.-X. Liu, L.-D. Zhang, S.-W. Wei and Y.-S. Duan,
{\it Localization and Mass Spectrum of Matters on Weyl Thick Branes},
JHEP {\bf 0808} (2008) 041.

\bibitem{LiuJCAP2009}
Y.-X. Liu, Z.-H. Zhao, S.-W. Wei and Y.-S. Duan,
{\it Bulk Matters on Symmetric and Asymmetric de Sitter Thick Branes},
JCAP {\bf 02} (2009) 003.

\bibitem{Guoetal2013}
H. Guo, A. Herrera-Aguilar, Y.-X. Liu, D. Malag\'on-Morej\'on and R.R. Mora-Luna,
{\it Localization of bulk matter fields on a pure de Sitter thick braneworld},
Phys. Rev.D {\bf 87} (2013) 095011.


\bibitem{Pomarol}
A. Pomarol,
 {\it Gauge bosons in a five-dimensional theory with localized gravity},
Phys. Lett. B {\bf 486} (2000) 153.

\bibitem{Oda}
 I. Oda,
{\it Bosonic fields in the string-like defect model},
Phys. Rev. D {\bf 62} (2000) 126009.

\bibitem{Oda:2001ss}
  I.~Oda,
{\it Locally localized gravity models in higher dimensions},
Phys.\ Rev.\ D {\bf 64} (2001) 026002.


\bibitem{GogMid1}
M. Gogberashvili and P. Midodashvili,
{\it Localization of gauge bosons in the 5D standing wave braneworld},
Phys. Lett. B {\bf 707} (2012) 169.

\bibitem{GogMid2}
M. Gogberashvili and P. Midodashvili,
{\it Gauge Fields in the 5D Gravity-Scalar Standing Wave Braneworld},
Europhys. Lett. {\bf 104} (2013) 50002.

\bibitem{OdaPLB2000113}
 I. Oda,
    {\it Localization of matters on a string-like defect},
    Phys. Lett. B {\bf 496} (2000) 113.

   \bibitem{HoffdaSilvaetal}
A.E.R. Chumbes, J.M. Hoff da Silva and M.B. Hott, { \it A model to localize gauge and tensor fields on thick branes},
 Phys. Rev. D {\bf 85} (2012) 085003.

\bibitem{XYYI}
   Z.-H. Zhao, Q.-Y. Xie, and Y. Zhong,
   {\it New localization method of U(1) gauge vector field on flat branes in (asymptotic) AdS5 spacetime},
   Class. Quantum Grav. {\bf32} (2015) 035020.
 \bibitem{MelfoPRD2006}
A. Melfo, N. Pantoja and J.D. Tempo,
{ \it Fermion localization on thick branes},
Phys. Rev. D {\bf 73} (2006) 044033.

\bibitem{Salvio:2007qx}
A.~Salvio and M.~Shaposhnikov,
{\it Chiral asymmetry from a 5D Higgs mechanism},
JHEP {\bf 0711}, 037 (2007).

\bibitem{Liu0708}
 Y.-X. Liu, L.-D. Zhang, L.-J. Zhang and Y.-S. Duan,
 { \it Fermions on thick branes in the background of Sine--Gordon kinks},
Phys. Rev. D {\bf 78} (2008) 065025.

\bibitem{Liu0907.0910}x
 Y.-X. Liu, J. Yang, Z.-H. Zhao, C.-E. Fu and Y.-S. Duan,
 { \it Fermion localization and resonances on {a de Sitter thick brane}},
 Phys. Rev. D {\bf 80} (2009)  065019.

\bibitem{Liusastb}
 Y.-X. Liu, C.-E. Fu, L. Zhao and Y.-S. Duan,
{\it Localization and mass spectra of fermions on
 symmetric and asymmetric thick branes},
Phys.  Rev. D {\bf 80}  (2009)  065020.

\bibitem{Koley2009}
 R. Koley, J. Mitra and S. SenGupta,
{\it Fermion localization in a generalized Randall-Sundrum model},
Phys. Rev. D {\bf 79} (2009) 041902(R).

\bibitem{liu_0909.2312}
Y.-X. Liu, H.-T. Li, Z.-H. Zhao, J.-X. Li and J.-R. Ren,
{\it Fermion resonances on multi-field thick branes},
JHEP {\bf 0910} (2009) 091.

\bibitem{Chumbes:2010xg}
A.~E.~R.~Chumbes, A.~E.~O.~Vasquez and M.~B.~Hott,
{\it Fermion localization on a split brane},
Phys.\ Rev.\ D {\bf 83}, 105010 (2011)

\bibitem{Castro:2010uj}
L.~B.~Castro and L.~A.~Meza,
{\it Fermion localization on branes with generalized dynamics},
Europhys.\ Lett.\  {\bf 102}, 21001 (2013).

\bibitem{YXuFWei}
Y.-X. Liu, Z.-G. Xu, F.-W. Chen, and S.-W. Wei,
{\it New localization mechanism of fermions on
braneworlds}, Phys. Rev. D {\bf 89} (2014) 086001.

\bibitem{HQuChu}
H. Guo, X. Qun-Ying, F. Chun-E , {\it Localization and quasilocalization of spin-1/2 fermion field on
two-field thick braneworld}, (2014) arXiv:1408.6155v1 [hep-th].

\bibitem{RAGAR}
R. Cartas-Fentevilla, Alberto Escalante, Gabriel Germ\'an, Alfredo
Herrera-Aguilar and Refugio Rigel Mora-Luna, {\it Coulomb's law
corrections and fermion field localization in a tachyonic de Sitter
thick braneworld }, arXiv:1412.8710 [hep-th].

\bibitem{Choudhury:2015wfa}
  S.~Choudhury, J.~Mitra and S.~SenGupta,
 {\it Fermion localization and flavour hierarchy in higher curvature spacetime}, arXiv:1503.07287 [hep-th].
\bibitem{A.Andrianov}
A. A. Andrianov, V. A. Andrianov, P. Giacconi, and R. Soldati, {\it
Domain wall generation by fermion self-interaction and light
particles}, JHEP {\bf07} (2003) 063.

\end{thebibliography}


\end{document}